# CODEX MUTABILIS: PRESERVING THE REASONS FOR CHANGES IN SCIENTIFIC NAMES

**Richard Littauer**

*Te Kura Mātai Pūkaha, Pūrorohiko, Te Herenga Waka School of Engineering and Computer Science, Victoria University of Wellington*
*Aotearoa New Zealand*
*richard.littauer@vuw.ac.nz*
*0000-0001-5428-7535*

**Jessamyn West**

*Flickr Foundation*
*USA*
jessamyn@gmail.com
0000-0001-6620-8910

***Abstract*** **– Digital preservation infrastructures often prioritize the stability of content and metadata. In taxonomy, species names are formed according to the Articles listed in the International Code of Zoological Nomenclature. The reasons for these changes are rarely recorded in detail or made machine-readable. This paper examines this preservation gap. Here, we cover issues in the Code by looking at approaches to recording nomenclatural changes related to the Kakapō *Strigops habroptilus.* As a potential solution, we present Codex Mutabilis, a digital journal with a publication model that documents ICZN-mandated name changes with full textual justification, persistent identifiers, and archival infrastructure. We argue that this model offers a blueprint for preserving interpretive metadata in the sciences.**

## I. INTRODUCTION

Digital preservation infrastructures which preserve changes to data as a primary artifact enable future researchers to determine the provenance and treatment of data. The International Code of Zoological Nomenclature (henceforth "the Code") [1] specifies how scientific names are formed but does not specify how to store the taxonomic decisions which influence changes in names. The result is instability, confusion, and duplicated work. These problems are not unique to the Code, but they do represent a unique challenge for digital preservation. Here, we discuss the issue of preserving interpretive metadata, and provide a case study involving the ICZN, using the newly formed journal Codex Mutabilis.

## II. BACKGROUND: INTERPRETIVE METADATA IS NOT PRESERVED

Some digital tools preserve editorial decisions as part of their tooling. For instance, Wikipedia users have to add a message to any changes they make on the platform. These messages are unstructured text fields with a timestamp. Likewise, the software program Git [2] contains commit messages which allow users to include a message with any associated change, and which contains the line-by-line changes for any edit to the data being stored. The Git commit messages are stored in a directed acyclical graph (DAG), which allows for any stage in a Git history to be examined at that stage. Wikipedia edits also allow users to see any version at any point in the editing history. This approach of storing changes with meaningful messages could also be applied elsewhere.

In taxonomic databases, the primary concerns for data preservation are the storage of scientific names, the authors who described those names, and the original publications where the names were made available according to the Code. The International Commission on Zoological Nomenclature (ICZN) publishes the Code, which is almost universally applied in zoological science. The Code stipulates how names are to be formed, what defines an “available” name – essentially, is the name valid according to the Code and does it match the taxonomic unit being described – and how to correct names which are formed incorrectly.

The Code says that species' names must agree in grammatical gender with the genus with which those species are combined. For instance, with the Kakapō *Strigops habroptilus*, the species name *habroptilus* must agree in gender with the generic name *Strigops*, which has historically been considered feminine. The Code has some rules around what must agree - for instance, adjectives agree with generic names, but nouns don’t. The determination of whether a species name is a noun or an adjective is determined by Article 31.2.2: "Where the author of a species-group name did not indicate whether he or she regarded it as a noun or as an adjective, and where it may be regarded as either and the evidence of usage is not decisive, it is to be treated as a noun in apposition to the name of its genus (the original spelling is to be retained, with gender ending unchanged; see Article 34.2.1)." [1]

Traditionally, taxonomists following this rule look at the original description, consult a Latin dictionary, and then present arguments in the literature. The ICZN very rarely makes rulings on certain species names overruling their own Code, and for the most part researchers are left to judge on their own whether a decision regarding a word was made correctly. These decisions are often made by taxonomic committees, which follow general tradition in their respective regions and taxa. They either debate the endings of words internally or are swayed by publications that deal with the particular case at hand. The taxonomies may add a short note in a changelog, but this is not necessitated by the Code.

For instance, *Strigops habroptilus* was first described by G.R. Gray [3], who used the final ending -*us*, which is masculine. In 2003, the Howard and Moore taxonomy [4] switched to *habroptila*. *Habroptilus* comes from ἁβρός *habros* “graceful” and πτιλον *ptilon* “feather” [5]. The editors of the checklist considered “graceful feathered” to be an adjective and therefore decided that it should be changed. There was no note in the taxonomy about this switch; the name alone was changed. Other taxonomies followed suit. Two decades later, Savage and Digby [5] published a retort to this change, noting that there is no way to know if “graceful feather” was a noun or an adjective, and that the original spelling was therefore correct according to Article 31.2.2. This publication was exceptional in that it considered only this particular species - normally, corrections for grammatical gender are done on an ad hoc basis by taxonomists while doing large revisions, but the charisma of a large ground parrot was enough to justify the special consideration.

Some taxonomies have since adopted the change, like BirdsNZ [6], Clements [7], and AviList [8]. The references they use to note this change are succinct. For instance, BirdsNZ noted: “We follow Savage & Digby (2023) in treating *Strigops* as masculine (contra ICZN 1955: 262 and Checklist Committee 2022), hence the species name should be *Strigops habroptilus* (not *S. habroptila*).” This note does not mention Article 31.2.2, nor the etymology of *habroptilus*. Other taxonomies may not include even this summary note for such changes.

Zoobank, which was set up by the ICZN, stores available names, but does not by default store changes under 31.2.2, or other changes mandated by the Code, such as the acts of First Revisers under Article 24. The lack of a standardized way to cite reasoning leads to confusion in the nomenclature. Any accepted change depends upon taxonomists reading all of the available literature for any given change, instead of depending upon a centralized body which preserves such decisions. The relevant literature may be locked behind paywalls, or may not exist, as appears to have been the case for the 2003 switch noted above in the Howard and Moore checklist [4] to *habroptila*.

The digital preservation of the reasoning behind names would help to clarify any changes, help explain editorial decisions by the editors of taxonomies, reduce the amount of time spent researching by taxonomists, and would serve as an interesting dataset in its own right. Any solution would have to provide:

- A list of Articles in the Code which are relevant for a change,

- A list of references of original documents used in the decision,
- A breakdown of the etymology of taxonomic names,
- An edited, peer-reviewed publication record of a change,
- A machine-readable interface,
- With version control,
- A persistent identifier, and
- Persistent storage of the data.

## III. Codex Mutabilis: A Model for Justification-Aware Preservation

Codex Mutabilis ("the changing book" in Latin) is a digital journal that was founded in order to provide a working solution to this problem, and which provides all of the above. It can be found at https://codexmutabilis.com.

The journal is built on a Jekyll website, served by GitHub actions through the GitHub Pages interface. GitHub is the largest site for code storage in the world, with over 150m developer accounts and almost half a billion projects [9]. It allows free hosting of websites using the GitHub Pages interface.

Codex Mutabilis has an editorial team which reviews any publications, which can be submitted through the open process provided by GitHub's open pull request system. This workflow is already being used by some taxonomies (such as AviList [8]) for their internal decision-making processes, and is also used by journals such as the Journal of Open Source Software [10]. The process allows peer-review, although a blind peer-review system has not yet been instituted.

Each publication to Codex Mutabilis is formulated as a Markdown file, which allows metadata which is machine readable. All work in Codex Mutabilis is stored in a Git repository, with associated version control for each change. Each published article has a standardized format for titles, references, and nomenclatural decisions. Arguments must follow a particular sequence.

Each article concerns only a single species, as each individual case must be judged to follow the Code. Articles are published on Zenodo and in Zoobank, which ensures a DOI. An example article can be seen in [11]. The use of tags allows for systematic metadata concerning nomenclatural decisions: for instance, with this article, the tag "substantive" was applied. The relevant Articles will also be tagged in future publications.

According to the Code, a published work must come registered with an ISSN. The ISSN for Codex Mutabilis is 3021-3592, and the work is currently being archived in the National Library of New Zealand. The stipulations for storage were that the work was not a personal blog, which was met by having an editorial staff of more than one, and that the work is relevant to New Zealand and is intended to be permanent and serial. The ISSN was procured in a couple of days, after these conditions were met.

Review publications are also accepted by Codex Mutabilis, for articles which discus a null change according to the Code. Any article explaining why a name is named the way it is named, with a history of the name, would be accepted for review. This helps ensure consistency and stability in the nomenclature, by reducing research time for contentious names and by serving as examples of how the Code is applied.

In short, Codex Mutabilis provides a publication record for each species considered, along with a persistent identifier and storage in three separate locations: GitHub, Zenodo, and the National Library. The publication record is machine readable, lists references, and is version-controlled through Git. Through tags and a standardized format, the articles store information about reasons applied according to the ICZN.

## IV. Conclusion

Codex Mutabilis is not perfect, but it is a first attempt at viewing the changes made according to the ICZN as relevant works worthy of digital preservation and explanation in their own right.

A larger schema for how this should be made. It could, for instance, involve flowcharts for which Articles of the ICZN were applied in sequence; integrate Linked Data formats or storage of acts in a DAG; or build a rigid lattice for framing bibliographic references for specific names. Other approaches would also be valid. Taxonomists could agree to provide structured information with their changes, and a metadata standard for nomenclatural decisions could be created by a taxonomic working group. The ICZN could also implement some of these changes in their next version of the code, which is forthcoming in the next few years.

This work could be applied to other domains. The Code is one of several major taxonomic codes mandating how scientific names are formed; for instance, botanists and virologists have their own codes. Further, the Code is not the only human system that could be worked in this way – a pattern like Codex Mutabilis could also be applied to legal or technical domains that depend upon strict protocols.

In the current system, changes according to the Code are published in the literature and then percolate through various taxonomies over time. The traditional scientific process for publication is slow, time-intensive, and demands high editorial commitment to well-formatted unstructured texts. These texts are often locked into HTML or PDF formats, making mining difficult. The model shown in Codex Mutabilis flips this model around, by emphasizing well-structured text that can be data-mined and archived, and by encouraging a short-form publication model that uses the traditional hallmarks of peer-review while also having persistent identifiers and storage in an institutional archive. This model potentially serves the Code better than a traditional publishing model, and it also serves to digitally preserve decisions in a way that makes them more accessible to researchers. It could also serve other Codes and other systems.

Codex Mutabilis shows how interpretive editorial acts, such as those pertaining to the ICZN Code, can be preserved cleanly and openly. More broadly, digital preservation should consider not only the data that must be preserved, but also the rationale behind those decisions.

## References

[1] International Commission on Zoological Nomenclature, *International Code of Zoological Nomenclature*, 4th ed. London, U.K.: International Trust for Zoological Nomenclature, 1999.

[2] L. Torvalds and J. Hamano, *Git: Fast Version Control System*, 2005. [Online]. Available: https://git-scm.com (accessed Apr. 15, 2025).

[3] G. R. Gray and D. W. D. W. Mitchell, *The genera of birds: comprising their generic characters, a notice of the habits of each genus, and an extensive list of species referred to their several genera*. London, Longman, Brown, Green, and Longmans, 1844-1849, 1844, vol. 2, https://www.biodiversitylibrary.org/bibliography/126497. [Online]. Available: https://www.biodiversitylibrary.org/item/219462 (accessed Apr. 15, 2025).

[4] E.C. Dickinson, Ed. *The Howard and Moore complete checklist of the birds of the world.* Third edition. Princeton University Press, Hawthorne, 2003.

[5] J.L. Savage and A. Digby, "Strigops habroptilus Gray, 1845 is the valid scientific name of the kākāpō (Aves, Strigopidae)," *Bull. Zoo. Nom.,* vol. 80, no. 1, pp. 112–115, 2023, doi: 10.21805/bzn.v80.a032

[6] C. M. Miskelly, N. J. Forsdick, R. L. Palma, N. J. Rawlence, and A. J. Tennyson, "Amendments to the 5th edition (2022) of the checklist of the birds of New Zealand," Notornis, vol. 71, no. 3, pp. 93–114, 2024, doi: 10.63172/871131edqjls

[7] Clements, "Updates and Corrections—October 2024 – Clements Checklist", birds.cornell.edu. https://www.birds.cornell.edu/clementschecklist/updates-and-corrections-october-2024/ (accessed Apr. 15, 2025).

[8] AviList Core Team, "AviList: The Global Avian Checklist, v2025," 2025, doi: 10.2173/avilist.v2025

[9] GitHub, "About GitHub", github.com. https://github.com/about (accessed Apr. 15, 2025).

[10] A. M. Smith, K. E. Niemeyer, D. S. Katz, L. A. Barba, G. Githinji, M. Gymrek, K. D. Huff, C. R. Madan, A. C. Mayes, K. M. Moerman *et al.*, "Journal of Open Source Software (JOSS): design and first-year review," *PeerJ Computer Science*, vol. 4, p. e147, 2018, doi: 10.7717/peerj-cs.147

[11] R. Littauer, "*Prunum boreale* (A. E. Verrill, 1884) is changed to *P. borealis*", Codex Mutabilis, Dec. 2024, doi: 10.5281/zenodo.14377608.

## Author Biographies

Richard Littauer is an avid birder, open science aficionado, and technologist. He serves on the BirdsNZ Checklist Committee and the AviList Bibliographic and Nomenclatural Committee. He founded Codex Mutabilis. He is also an organizer for SustainOSS and CURIOSS.

Jessamyn West is a librarian and technologist living in rural Vermont. She has worked at the intersection of libraries, technology, and politics since 2003. Her professional blog, Librarian.net, was one of the first librarian blogs.